\documentstyle[sprocl]{article}

\input{psfig}

\def\Journal#1#2#3#4{{#1} {\bf #2}, #3 (#4)}
\def\Cite#1{$^{[}$\cite{#1}$^{]}$}

\def\NIMA{{\em Nucl. Instrum. Methods} A}

\def\PLB{{\em Phys. Lett.}  B}
\def\PRL{\em Phys. Rev. Lett.}
\def\PRC{{\em Phys. Rev.} C}

\def\JPG{{\em J. Phys.} G}

\def\be{\begin{equation}}
\def\ee{\end{equation}}
\def\bea{\begin{eqnarray}}
\def\eea{\end{eqnarray}}

\begin{document}

\title{Production of Light (anti-)Nuclei with E864 at the AGS}
\author{Zhangbu Xu for the E864 Collaboration}
\address{Physics Department, Yale University\\ CT 06520, USA\\E-mail: 
xzb@hepmail.physics.yale.edu} 
\maketitle\abstracts{Light nuclei can be produced in the central 
reaction zone via coalescence in relativistic heavy ion collisions. 
E864 at BNL has measured the production of ten stable light nuclei 
with nuclear number of $A=1$ to $A=7$ at rapidity $y\simeq1.9$ and 
$p_{T}/A\leq300MeV/c$. Data were taken with a Au beam of momentum of 
$11.5$ A $GeV/c$ on a Pb or Pt target with different experimental settings.
The invariant yields show a striking exponential dependence on nuclear number
over ten orders of magnitudes with a penalty factor of about 50 per 
additional nucleon. This penalty factor is used to estimate the strange 
quark matter (strangelet) production in the baryon rich and strangeness 
enhanced environment. The measurements of the production of antiproton, 
antideuteron, hypernuclei ($^{3}_{\Lambda}H$,$^{4}_{\Lambda}H$), 
and strongly unstable nuclear states ($^{4}H$,$^{5}Li$, $^{5}Li^{*}$, 
$^{5}He$) are presented as well. A model of local thermal equilibrium with 
radial flow at the kinetic freeze-out with a temperature of 
$T=112\pm10MeV$, chemical potential of $\mu_{B}=503\pm20MeV$ and flow 
velocity of about $\sqrt{V_{\perp}^{2}}\simeq0.5c$ 
seems to be able to describe the data in 
the gross scale with the exceptions of the production of antihyperons and 
hypernuclei. The large antihyperon production and the extra penalty for 
hypernuclei production are quite surprising.}
{\bf 1. Introduction}

    Relativistic heavy ion collisions may create high energy density and
high baryon density in the reaction zone. Light nuclei can be produced
by the recombination of created or stopped nucleons\Cite{nagle}.
This recombination process is called coalescence. Coalescence of nuclear 
clusters can be characterized by the penalty factor for a nucleon added 
to the nuclear cluster. This idea can be extended to hypernuclei to 
calculate the strangeness penalty 
factor. Both baryon and strangeness penalty factor are 
useful to estimate the production rate of strange quark matter in terms 
of coalescence or thermal production. 
Since the probability of coalescence of a particular nuclear system
(d, $^{3}He$, etc.) depends on the properties of the hadronic system formed
as a result of the collision, the study of the coalescence process is useful
in elucidating those properties. For example, in a coalescence model, the
coalescence probability depends on the temperature, baryon chemical potential
(essentially the baryon density), and the size of the system, as well as the
statistical weight (degeneracy) of the coalesced nucleus. 
The data reported shows evidence that the probability
may also depend on the binding energy of the coalesced nucleus at the level of 
about 6MeV. Systematic study of the production of light nuclei is limited
by their low production rates in relativistic heavy ion collisions.
E864 is the only experiment which is able to measure the production of charged
nuclei with $A>4$ produced in the central reaction zone. However, the
nature of the coalescence process tells us that the higher the baryon number,
the more sensitive the production rate is to the system's configuration. 
For example, the inverse slope of the transverse momentum distribution 
(effective temperature) of ``heavy'' light nuclei is very sensitive to the 
density profile and radial flow profile. Production of antinuclei will provide 
additional information about the equation of state of the system and the 
state of chemical or kinetic equilibrium. 

In this report, the production of p, n, d, $^{3}He$,
t, $^{4}He$, $^{6}He$, $^{6}Li$, $^{7}Li$ and $^{7}Be$ around rapidity
$y\simeq1.9$ and transverse momentum of $p_{T}/A\leq300MeV/c$ in 10\% most
central Au+Pt(Pb) collisions measured by E864\Cite{nim} at BNL is
presented together with the production of ${\bar p}$, ${\bar d}$, $^{5}He$, 
$^{5}Li$, $^{4}H$, $^{3}_{\Lambda}H$ and the production limits on 
$^{5}Li^{*}$ and $^{4}_{\Lambda}H$.
These measurements have significant impact on the strange
quark matter\Cite{jaffe} search by several experiments in relativistic
heavy ion collisions. They also provide information about the thermal
equilibrium of the system and the detailed process of coalescence. 
The richness of the data in rapidity and centrality\Cite{nigel} will not be 
discussed here.\\
{\bf 2. E864 Apparatus}

  E864\Cite{nim} is an open-geometry and
high-data rate apparatus designed to search for new forms of matter. It has
large rapidity and transverse momentum coverage. The tracking system 
consists of 2 dipole magnets for rigidity measurement and particle selection,
3 hodoscopes and 2 straw tube chambers measuring velocity, 
charge and position. 
The mass resolution is about 3\% in the region of interest in high field 
settings.There is an additional mass measurement from hadronic calorimeter, 
which provides good energy and TOF information. These signals are used for 
a high mass level two trigger as well. More detailed descriptions of 
the apparatus can be found in other publications\Cite{nim,nigel,xzb,evan}.

First, we can measure single-particle spectra of particles, such as
$^{6}He$, $^{7}Li$, $^{7}Be$, n, ${\bar p}$, $K^{\pm}$ and $\pi^{\pm}$.  
Because E864 is open-geometry configuration, we have multiple particles in
one event. We can also do two-particle correlation to reconstruct the invariant
mass of their decay parent. We use event-mixing to subtract the background.
Since we do not identify the decay vertex, strong decays and weak decays
appear the same technically. 
Last but not the least, from the large sample of events, we can 
look for rare products (${\bar d}$) and search for new forms of matter 
such as strange quark matter.\\
{\bf 3. Results and Discussions}

 In this section, we will present the results of the single-particle 
spectra and apply a simplified coalescence model assuming local thermal 
equilibrium to describe the data in the gross scale. Systematic errors 
of the temperature and chemical potential extracted from this 
model-dependent simplification\Cite{heinz} will be discussed at the end. 
Radial flow is then studied using the temperature obtained and the inverse 
slope parameters from the transverse spectra of the $K^{+}$, p, 
n, d, $^{3}He$ and $^{4}He$(preliminary). 
Then we will show the results on 
the production of the hypernuclei and address the strangeness penalty factor. 
At the end, we will present the limits on strangelet search and discuss 
the significance of the limits in the context of the coalescence model. 
Particularly, we will discuss the limits on H-d (bound state of H-dibaryon 
and deuteron) search. 

\begin{figure}
\centering\psfig{figure=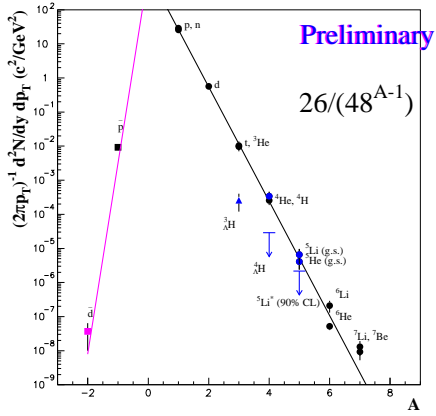,height=2.2in}
\caption{Eighteen hadrons measured by E864. Stable light 
nuclei\protect\Cite{xzb} are selected from \(y=1.9\) and 
\(p_{T}/A<300MeV/c\). ${\bar p}$ is at $y=1.9$ and 
$p_{T}\simeq0$\protect\Cite{pbar}. ${\bar d}$ is at $y=1.9$ and 
$0<p_{T}<1GeV/c$(preliminary). Hypernuclei are from $1.6<y<2.6$ with 
correction of detector's acceptance(preliminary). Strongly unstable 
nuclei are from $y=1.9$ and $p_{T}{}_{\sim}^{<}2GeV/c$(preliminary). 
Lines are fits to the data assuming local thermal equilibrium
(work in progress). See text for details.\label{fig:all18hadrons}}
\end{figure}

\begin{figure}
\begin{minipage}[b]{0.46\linewidth}
\centering\psfig{figure=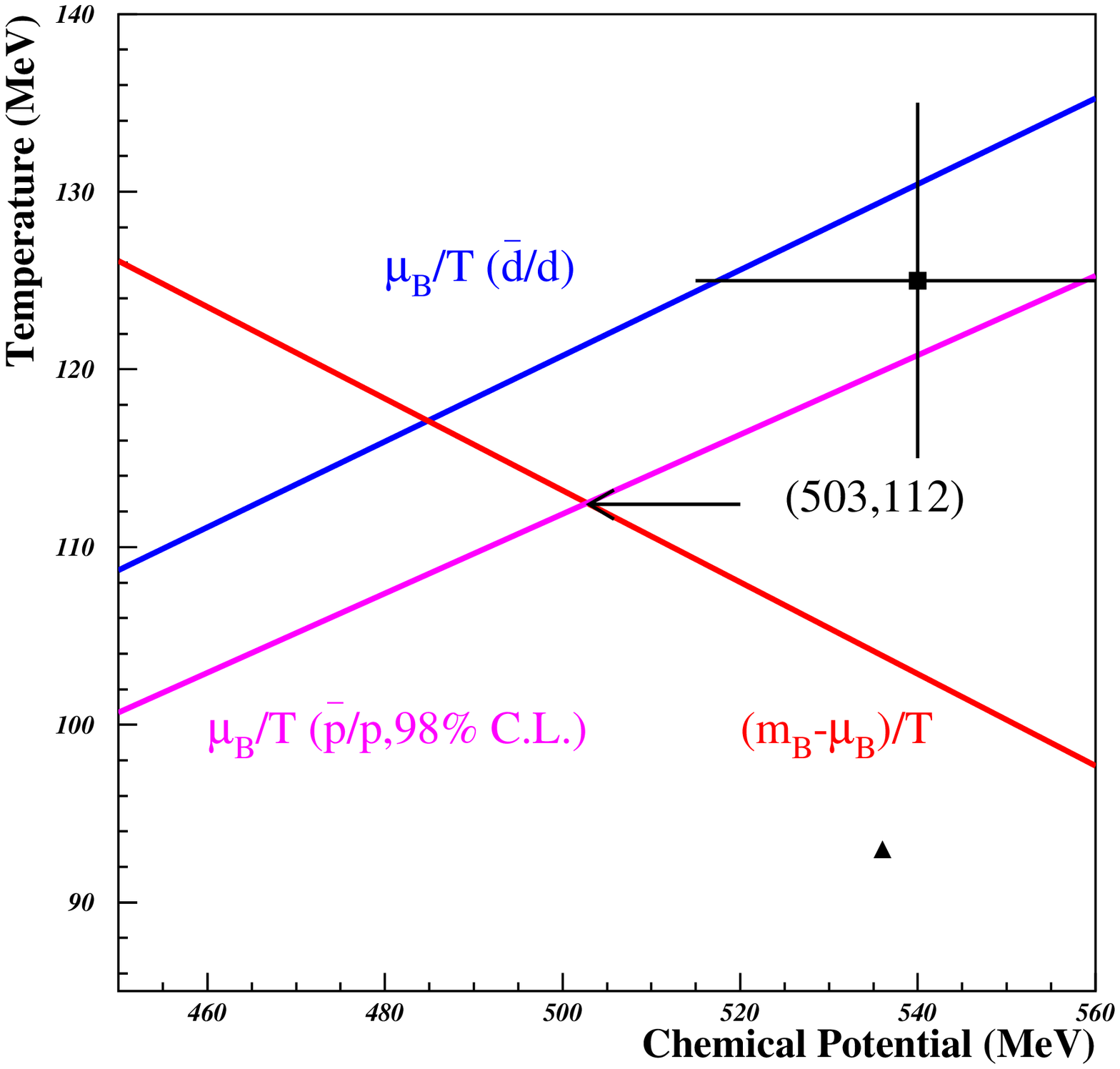,width=\linewidth}
\caption{Temperature vs. chemical potential. The square and triangle points 
are from \protect\Cite{braun,heinz}. $T=112\pm10MeV$, $\mu_{B}=503\pm20MeV$ 
with estimated uncertainty. Errors are not shown. 
See text for details}\label{fig:mu_t}
\end{minipage}\hfill
\begin{minipage}[b]{0.46\linewidth}
\centering\psfig{figure=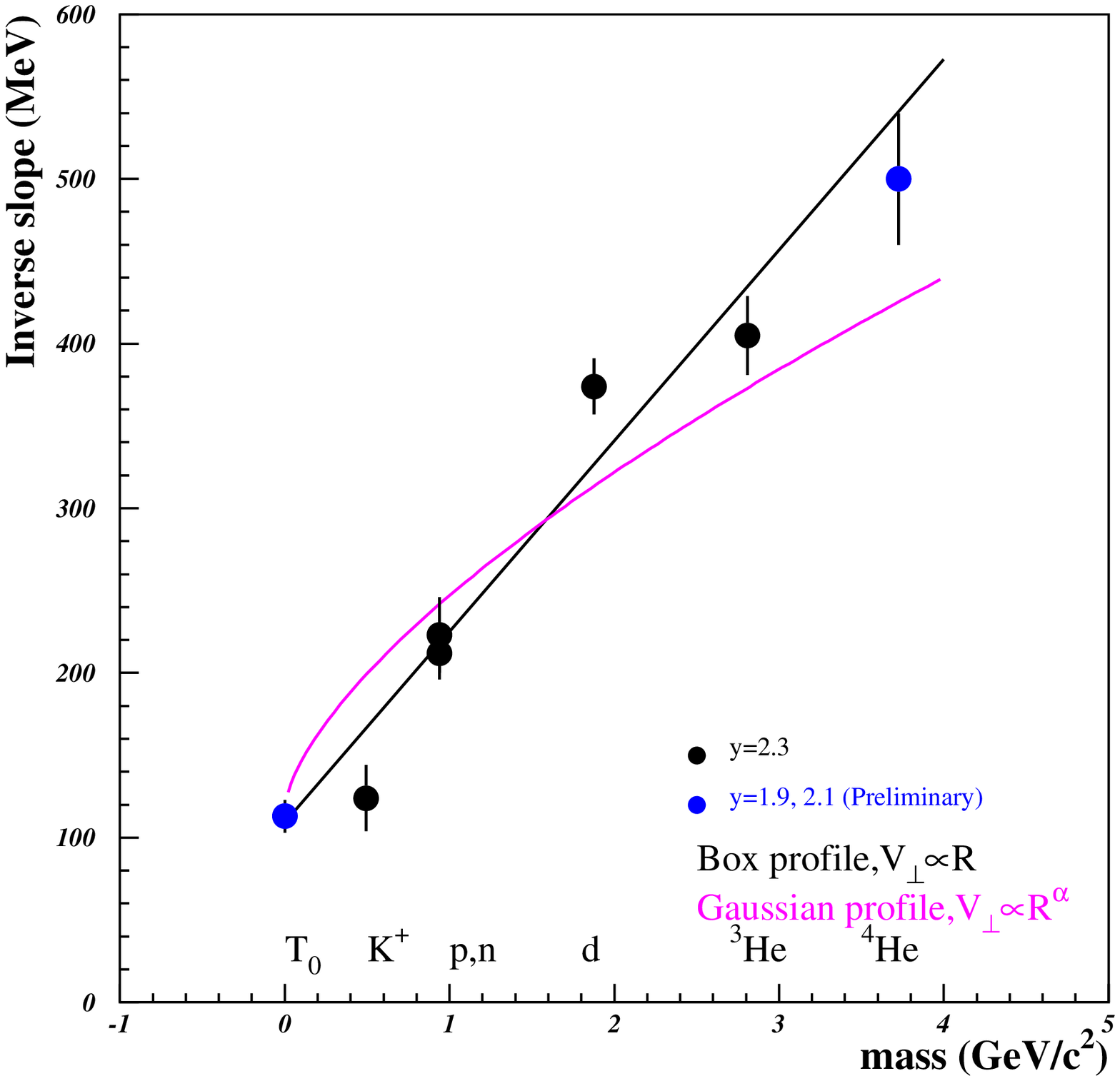,width=\linewidth}
\caption{Radial flow from transverse momentum spectra. $T$ and inverse slope 
of $^{4}He$ are at $y=1.9$ and $y=2.1$ respectively which have higher 
temperature and stronger flow. 
The curved line is to guide the eye.}\label{fig:flow}
\end{minipage}
\end{figure}

Figure~\ref{fig:all18hadrons} shows the invariant yields of 18 hadrons 
we have measured in a selected rapidity ($y$) and transverse momentum 
($p_{T}$) bin. We observe a striking exponential dependence of light nuclei 
production up to $A=7$ over ten orders of magnitudes in the invariant yield. 
Here, we will briefly present the results on ${\bar d}$ production as an 
example.
Invariant yields are calculated and presented in terms of
$d^{2}N/(2\pi p_{T}dp_{T}dy)$ in units of $GeV^{-2}c^{2}$ per central 
collision. 
In case of ${\bar d}$, we have a sampled collection of $1.4\times10^{10}$ 
events of 10\% most central interactions. A signal of $17.6\pm7.5$ in the  
rapidity range of $1.8<y<2.2$ and $4.6\pm3.3$ in $1.4<y<1.8$ above background 
is observed. The yields are determined to be $3.5\pm1.5
(stat)^{+0.9}_{-0.5}(sys)\times10^{-8}GeV^{-2}c^{2}$ and 
$3.7\pm2.7(stat)^{+1.4}_{-1.5}(sys)\times10^{-8}GeV^{-2}c^{2}$ 
respectively\Cite{gene}. These measurements suggest that the coalescence 
parameters $B_{2}$ and ${\bar B_{2}}$ are consistent with each 
other\Cite{gene}. In this report, we will assume that antinuclei and nuclei 
have the same temperature ($T$) and chemical potential ($|\mu_{B}|$) at the 
kinetic freeze-out.\\
We fit the production of the ten stable light nuclei with an exponential as 
a function of nuclear (baryon) number. The best fit\Cite{xzb} is 
$26/48^{A-1}$. If we take our ${\bar p}$ and correct it for the antihyperon 
feeddown\Cite{pbar} at 98\% confidence level (CL)in $y=1.9$ near 
$p_{T}\simeq0$, we have the fugacity\Cite{heinz}(Eq.6.6) of 
$\lambda^{2} =$${\exp({{2\mu_{B}}\over T})}$$= p/{\bar p} = 23.3/0.0038.$
The fugacity can also be obtained from $d/{\bar d}$: 
$\lambda^{\prime 4} = d/{\bar d}=0.56/(3.5\times10^{-8})$.
The difference between $\lambda$ and $\lambda^{\prime}$ comes from the 
difference of the isospin abundance or $n/p$ ratio in a thermal distribution. 
From our unique measurement\Cite{evan} of $n/p=1.19\pm0.08$, we have 
$\lambda^{\prime}=1.1\lambda$.
For simplicity, the average fugacity or chemical potential are taken as:
$\lambda_{B}=\sqrt{\lambda\lambda^{\prime}}$,
and 
$\mu_{B} = (\mu_{p}+\mu_{n})/2$. In fact, the fugacity from $p/{\bar p}$ 
at 98\% CL is within the error of that from $d/{\bar d}$. 
If we consider the penalty factor coming 
from the possibility of the population of the state as in a thermal system
and neglect all the other details(flow has built up before the coalescence 
process and there is not additional momentum redistribution of the cluster 
after-burner), we have 
$\exp((m_{N}-\mu_{B})/T)\simeq48$ where $m_{N}$ is the nucleon mass. In 
Figure~\ref{fig:mu_t}, we plot the three equations above in the 
phase space of chemical potential and temperature. From the crossing point 
of the lines from penalty factor and fugacity from $p/{\bar p}$, we get 
$T=112\pm10MeV$ and $\mu_{B}=503\pm20MeV$ 
(errors are estimates,work in progress). This is 
quite close to other measurements\Cite{braun,heinz}.\\
In the non-relativistic limit which for the case of nuclei is a good 
approximation, the kinetic energy 
can be separated into thermal energy and flow energy as 
$T_{eff} = T+{1\over2}m{\bar V}_{\perp}^{2}$ in two-dimension radial flow. 
Figure~\ref{fig:flow} is the effective temperature vs. mass up to 
$A=4$ with $T=112MeV$ at $m=0$. Data of p, n, d, 
$^{3}He$ are at $y=2.3$ where E864 has its best coverage in transverse 
momentum. For the flow at $y=2.3$, we have to scale 
T ($y=1.9$) and inverse slope parameter of $A=4$($y=2.1$) down by 10\% to 
20\% which was not done in the plot. 
Therefore, the flow velocity we get probably will be an overestimate. 
The fit is $T_{eff}=112+117\times A (MeV)$ with a mean velocity 
 $\sqrt{{\bar V}_{\perp}^{2}}\simeq0.5$. \\
Now we can use the flow and rapidity distribution measured to 
estimate the systematic error introduced by the model-dependent 
assumption\Cite{heinz} that the system is in local thermal equilibrium. 
We can compare the difference between the total yield and the 
yield in one specific rapidity and momentum. When the parametrization 
of the correction factor of ${A}^{\chi}$ from \Cite{heinz} 
Eq.6.10 is used, the limits of penalty between 39 and 72 are obtained for 
$1>\chi>-{1\over2}$. 
If we assume that the temperature distribution in rapidity is a Gaussian 
with $\sigma_{T}=1.1$ from our neutron measurements\Cite{evan} and the 
effective temperature scales as $T_{eff}\simeq1.0+A$ with Boltzmann 
distribution, then 
$T_{eff}(y)\propto(1.0+A)\times\exp{(-(y-y_{cm})^{2}/(2\sigma_{T}^{2}))}$.
We have characterized 
the yield in rapidity at $p_{T}\simeq0$ by the concavity\Cite{nigel} $b/a$ 
and extrapolated to other nuclei\Cite{xzb} with limited data points. 
We obtain $b/a=0.14, 0.92, 2.8, 2.4, 3.2, 7.7$ for p,d, $^{3}He$, t, $^{4}He$, 
$^{6}He$ respectively\Cite{nigel,xzb}. The penalty factor becomes 25 when 
we integrate over the transverse momentum and rapidity of 
$-0.6<y-y_{cm}<0.6$(projectile and target regions are not included to avoid 
fragmentations, 
$dN/dy\propto T_{eff}(T_{eff}+m){{dN}\over{dyp_{T}dp_{T}}}|_{p_{T}=0}$), 
which should probably be taken as the 
absolute lower limit because the integration assumes non-correlation 
between momentum and space in the extrapolation to high $p_{T}$ for high mass 
which is in contradict with the flow phenomenon; in addition, 
there is indication that the effective temperature does not 
scale linearly with mass and might saturate at high masses as shown in 
Figure~\ref{fig:flow}. In conclusion, the penalty factor is about 
$48^{+24}_{-10}$ with uncertainty dominated by model-dependent estimate.

  E864 has a sampled collection of about $9\times10^{10}$ events of 
10\% most central interactions for hypernuclei studies. We have analyzed 
$2/3$ of the data and found a $^{3}_{\Lambda}H$ signal at $1.8$ sigma level 
above the background after mixed-event background subtraction\Cite{batsouli}. 
The invariant yield is $2.6\pm1.4\times10^{-4}Gev^{-2}c^{2}$ over 
$1.6<y<2.6$ and $0<p_{T}<2GeV/c$. The strangeness penalty factor is 
$\lambda_{S}={}^{3}_{\Lambda}H/{}^{3}He=0.036\pm0.019$ with an extra penalty 
of ${{{}^{3}_{\Lambda}H\times p}\over{{}^{3}He\times\Lambda}}=0.17\pm0.09$
For the $^{4}_{\Lambda}H$, we do not observe any significant signal above 
background and we set a upper limit at 90\% CL of 
$0.29\times10^{-4}Gev^{-2}c^{2}$. When the degeneracy of $\Sigma(2J+1)=4$ is 
taken into account, we have 
$\lambda_{S}={}^{4}_{\Lambda}H/4/{}^{4}He<0.032$ which is consistent with 
the $^{3}_{\Lambda}H$ case. This extra penalty may come from the quantum 
mechanical correction\Cite{heinz}.
From this, we can write the cluster 
production\Cite{xzb} as $157/50^{A-1}/30^{|S|}$. The penalty for antimatter 
coalescing is much larger ($\simeq10^{6}$).

We have searched for strangelets (small strange quark matter) with charge of 
0,$\pm1,\pm2,\pm3$ with null result; limits are set between $10^{-9}$ to 
$10^{-8}$ per central collision over a large mass range. We can discuss the 
sensitivity in terms of a coalescence mechanism and write the 
sensitivity \Cite{xzb} as $A+0.87|S|{}^{<}_{\sim}7.1$ with the newly 
measured $|\lambda_{S}|$. 
For example, we set a limit of $0.9\times10^{-7}$ for long-lived H-d state 
which has been predicted to be produced at a level of 
$10^{-5}\rightarrow10^{-7}$ by C. Dover or 
$157/50^{3}/30^{2}=1.4\times10^{-6}$ from our measurements\Cite{xzb} if it 
exists. 

In summary, we have measured the production of 18 hadrons in 
relativistic heavy ion collisions at the AGS energies. Information of the 
colliding system, such as temperature, chemical potential, flow and 
cluster production rate can be extracted from these measurements.
Z. Xu would like to thank Dr. U. Heinz and Dr. J. Stachel for valuable 
discussions. We gratefully acknowledge the excellent support of the AGS staff. 
This work was supported in part by grants from the U.S. Department of 
Energy's High Energy and Nuclear Physics Divisions, and the U.S. National 
Science Foundation.\\

\end{document}